\documentstyle[12pt,aaspp4]{article}
\begin{document}

\newcommand{\up}[1]{\ifmmode^{\rm #1}\else$^{\rm #1}$\fi}
\newcommand{\zdot}{\makebox[0pt][l]{.}}
\newcommand{\upd}{\up{d}}
\newcommand{\uph}{\up{h}}
\newcommand{\upm}{\up{m}}
\newcommand{\ups}{\up{s}}
\newcommand{\arcd}{\ifmmode^{\circ}\else$^{\circ}$\fi}
\newcommand{\arcm}{\ifmmode{'}\else$'$\fi}
\newcommand{\arcs}{\ifmmode{''}\else$''$\fi}

\title{
The Araucaria Project: The Distance to the Sculptor Group Galaxy 
NGC 247 from Cepheid Variables Discovered in a Wide-Field 
Imaging Survey
\footnote{Based on  observations obtained with the 1.3~m
telescope at the Las Campanas Observatory, the 2.2 m ESO/MPI telescope at the European Southern Observatory
for Large Programme 171.D-0004, and the
4.0 m Blanco telescope at the Cerro Tololo Inter-American Observatory. This work is part
of the PhD thesis of AGV.}
}

\author{Alejandro Garc\'{\i}a-Varela}
\affil{Universidad de Concepci{\'o}n, Departamento de Fisica, Astronomy
Group,Casilla 160-C,
Concepci{\'o}n, Chile}
\authoremail{agarcia@astro-udec.cl}
\author{Grzegorz Pietrzy{\'n}ski}
\affil{Universidad de Concepci{\'o}n, Departamento de Fisica, Astronomy
Group,
Casilla 160-C,
Concepci{\'o}n, Chile}
\affil{Warsaw University Observatory, Al. Ujazdowskie 4,00-478, Warsaw,
Poland}
\authoremail{pietrzyn@astrouw.edu.pl}
\author{Wolfgang Gieren}
\affil{Universidad de Concepci{\'o}n, Departamento de Fisica, Astronomy Group, 
Casilla 160-C, 
Concepci{\'o}n, Chile}
\authoremail{wgieren@astro-udec.cl}
\author{Andrzej Udalski}
\affil{Warsaw University Observatory, Aleje Ujazdowskie 4, PL-00-478,
Warsaw,Poland}
\authoremail{udalski@astrouw.edu.pl}
\author{Igor Soszy{\'n}ski}
\affil{Universidad de Concepci{\'o}n, Departamento de Fisica, Astronomy
Group, Casilla 160-C, Concepci{\'o}n, Chile}
\affil{Warsaw University Observatory, Aleje Ujazdowskie 4,
PL-00-478,Warsaw, Poland}
\authoremail{soszynsk@astrouw.edu.pl}
\author{Alistair Walker}
\affil{Cerro Tololo Inter-American Observatory, Casilla 603, La Serena Chile}
\authoremail{awalker@ctio.noao.edu}
\author{Fabio Bresolin}
\affil{Institute for Astronomy, University of Hawaii at Manoa, 2680 Woodlawn 
Drive, 
Honolulu HI 96822, USA}
\authoremail{bresolin@ifa.hawaii.edu}
\author{Rolf-Peter Kudritzki}
\affil{Institute for Astronomy, University of Hawaii at Manoa, 2680 Woodlawn 
Drive, Honolulu HI 96822, USA}
\authoremail{kud@ifa.hawaii.edu}
\author{Olaf  Szewczyk}
\affil{Universidad de Concepci{\'o}n, Departamento de Fisica, Astronomy Group,
Casilla 160-C, Concepci{\'o}n, Chile}
\affil{Warsaw University Observatory, Aleje Ujazdowskie 4, PL-00-478,
Warsaw,Poland}
\authoremail{szewczyk@astro-udec.cl}
\author{Micha{\l} Szyma{\'n}ski}
\affil{Warsaw University Observatory, Aleje Ujazdowskie 4, PL-00-478,
Warsaw, Poland}
\authoremail{msz@astrouw.edu.pl}
\author{Marcin Kubiak}
\affil{Warsaw University Observatory, Aleje Ujazdowskie 4, PL-00-478,
Warsaw, Poland}
\authoremail{mk@astrouw.edu.pl}
\author{{\L}ukasz Wyrzykowski}
\affil{Warsaw University Observatory, Aleje Ujazdowskie 4, PL-00-478,
Warsaw,Poland}
\affil{Institute of Astronomy, University of Cambridge, Madingley Road, CB3 0HA, UK}
\authoremail{wyrzykow@astrouw.edu.pl}

\begin{abstract}
We report on the discovery of a Cepheid population in the Sculptor Group 
spiral galaxy NGC 247 for the first time. 
On the basis of wide-field images collected in photometric surveys in V and
I bands which were conducted with three different telescopes and cameras, 
23 Cepheid variables were
discovered with periods ranging from 17 to 131 days. We have constructed
the period-luminosity relations from these data and obtain distance moduli to
NGC 247 of 28.20 $\pm$ 0.05 mag (internal error) in V, 28.04 $\pm$ 0.06
mag in I, and 27.80 $\pm$ 0.09 mag in the reddening-independent Wesenheit
index. From our optical data we have determined the total mean reddening of the
Cepheids in NGC 247 as E(B-V)=0.13 mag, which brings the true distance modulus
determinations from the V and I bands into excellent agreement with the
distance determination in the Wesenheit index. The best 
estimate for the true distance modulus of NGC 247 from our optical Cepheid photometry 
is 27.80 $\pm$0.09 (internal error) $\pm$ 0.09 mag (systematic error) which is in
excellent agreement with other recent distance determinations for NGC 247
from the Tip of the Red Giant branch method, and from the Tully-Fisher relation.
The distance
for NGC 247 places this galaxy at twice the distance of two other Sculptor
Group galaxies, NGC 300 and NGC 55, yielding supporting evidence for the
filament-like structure of this group of galaxies.
The reported distance value
is tied to an assumed LMC distance modulus of 18.50 mag. \\
\end{abstract}


\section{Introduction}
In order to improve the extragalactic distance scale using stellar
standard candles observable in nearby galaxies, the Araucaria
Project\footnote{http://ezzelino.ifa.hawaii.edu/$\sim$bresolin/Araucaria/index.html}
(Gieren et al. 2001) has been investigating the most promising stellar
distance indicators: Cepheid variables, RR Lyr{\ae} stars, red clump
stars, blue supergiants  and the tip of the red giant branch (TRGB).
The main  goal of our project is to obtain an accurate
calibration, better than 5\%, of these primary distance indicators as a
function of the environmental properties of the host galaxy: metallicity,
age and star formation history. To achieve this goal, the Araucaria
Project studies galaxies of different morphological types, ages and
metallicities in the Local Group,
and in the more distant Sculptor Group which contains a number of
spiral galaxies easily accesible from the southern hemisphere. 

One of the Araucaria target galaxies is the spiral
galaxy NGC 247 (Fig. 1), member of the Sculptor Group (Jerjen et al. 1998;
Karachentsev 2005). This spiral classified as SAB(s)d in the
NASA/IPAC Extragalactic Database  and located in the constellation of
Cetus near the Galactic South Pole and is far from the Galactic Plane,
as indicated by its galactic coordinates  $\ell = 113.95 \deg$, $b =-83.56 \deg$.
In the following, we give a brief overview about previous studies of
NGC 247 which have measured the most important properties of this galaxy.\\
The first authors conducting  a  photoelectric study of bright stars in
NGC 247 in the U, B, V, R and I bands were Alcaino \& Liller (1984) who
provided a fundamental reference for the subsequent photometric
calibration of resolved stellar populations of this galaxy.\\
>From Schmidt plates, Carignan (1985) studied the surface brightness of NGC 247
and computed a distance modulus of 27.01 mag for the galaxy. It was also possible to
establish that NGC 247 has an inclination of 75.4 $\deg$, indicating the
difficulty involved in the study of its stellar populations due to the
problems of crowding and blending accentuated by this high inclination.

Through a study of the properties of the neutral hydrogen and the mass
distribution using VLA data, Carignan \& Puche (1990) established that the
HI disk of NGC 247 is relatively limited in extension. Based on their distance of 2.52 Mpc
(Carignan 1985), these authors determined the total mass in HI as (8.0$
\pm$ 0.3) $\times 10^{8} M_{\odot}$, which corresponds to a
mass-luminosity ratio of $M_{H_I}/L_B=$0.34.

Str\"assle et al. (1999) made a study in the X-ray energy range between
0.25 and 2 Kev using  ROSAT data. They found that the mass of the hot gas of
NGC 247, which emits mainly in the band of 0.25 Kev, is 0.7 $\times 10^{8}
M_{\odot}$. Taking the model for dwarf galaxies of Burlak (1996), which
integrates three components of the galaxy: stellar disk, gaseous disk and
dark halo, these authors found that the total mass of NGC 247 is 7.5
$\times 10^{10} M_{\odot}$, a value consistent with their measurements of
the rotation curve for this galaxy. Based on these data, Str\"assle et al.
(1999) found a mass-luminosity ratio of $M/L_x$=28.4, which is in
agreement with determinations of the same ratio for other Sculptor
galaxies.

Using the MOSAIC II camera at the 4.0 m Blanco
telescope at Cerro Tololo, Olsen et al. (2004) conducted a modern study of
globular clusters in the Sculptor galaxies. These authors were able to detect
a very small population of three globular clusters in NGC 247.

Karachentsev et al. (2006) were the first to measure the distance to NGC 247
from the Tully-Fisher method and obtained a value of 27.87 $\pm$ 0.21 mag,
in significant disagreement with the earlier, shorter distance value obtained by
Carignan.
In their infrared study of the central regions of nearby galaxies, Davidge
\&  Courteau (2002) had adopted a distance modulus of  27.3 mag for NGC 247.
The most recent study reporting a measurement of the distance of NGC 247
from a stellar population was performed by Davidge (2006). Based on the I-band
magnitude of the TRGB, he obtained a distance modulus of 27.9 $\pm$
0.1 mag , which implies a distance of 3.80 $\pm$ 0.18 Mpc, making this
galaxy one of the most distant targets of the Araucaria Project. Taking
into account the inclination of the NGC 247 disk, Davidge stressed that
variable internal extinction is expected, since the galaxy inclination
favors the possibility that the line of sight passes through dust clouds
more than once.
Freedman et al. (1988), and Catanzarite et al. (1994) were the first to
mention the discovery of a number of Cepheid variables in NGC 247. They reported that 
twelve Cepheids
had been found from  optical surveys in B, V, R and I bands conducted at
CTIO and LCO. Unfortunately, a catalog and/or photometric data for these Cepheids 
were never published.

This paper is the first study which reports  the discovery of a sizeable Cepheid
population in NGC 247 in a verificable catalog. The paper is structured as
follows: details of the observations, data reductions, and calibrations
are described in \S 2. The catalog of photometric properties of the 23
Cepheid variables in NGC 247 we have discovered is presented in \S 3. The determination
of the distance of NGC 247 from the observed period-luminosity relations in the V and I bands, and in the
reddening-independent Wesenheit index is the subject of \S 4. We discuss our results  
in \S 5 and summarize our conclusions in \S6.

\section{Observations,  Reductions and Calibrations}
NGC 247 was  observed with the 1.3 m Warsaw telescope at LCO, the 2.2 m
MPG/ESO telescope at La Silla Observatory and the 4.0 m Blanco telescope at CTIO.
Each telescope was equipped with a mosaic 8k $\times$ 8k detector with a
field of view (FOV)  and a scale factor (SF) as indicated in the Table 1.
For more instrumental details on these cameras, the reader is referred to
the camera Web
sites.\footnote{http://ogle.astrouw.edu.pl/~ogle/index.html,
http://www.ls.eso.org/lasilla/sciops/2p2/E2p2M/WFI/ and
http://www.ctio.noao.edu/mosaic/}

The observations of NGC 247 with the 1.3 m Warsaw telescope  were conducted in
service mode. The collection of data began on 2003 August 5, and ended on
2004 December 18. We obtained 40 epochs in the V-band and 38
epochs in the I band. With the 2.2 m telescope at La Silla, service mode observations 
in the V band were obtained for 19 epochs between 2003
June 23 and 2005 November 6. With the CTIO 4.0 m telescope, we obtained 
data in the V and I bands during 9 epochs.
These observations were conducted in visitor mode between 2002 October 30 and  2005 January
17. Table 2 summarizes the information about the observations for each
telescope.

For all images, the  debiasing and flat-fielding
reductions were done with the IRAF\footnote{ IRAF is distributed by the
National Optical Astronomy Observatory, which is operated by the
Association of Universities for Research in Astronomy, Inc., under
cooperative agreement with the NSF.} package. Then, point-spread function
photometry was obtained for all stars in the same way described by
Pietrzy\'nski et al. (2002a). Independently, data taken with the 1.3 m
Warsaw telescope were reduced with the OGLE-III pipeline based on the
image subtraction technique (Udalski 2003; Wo\'zniak 2000). \\
In order to perform an accurate calibration of our instrumental photometry
obtained with the 1.3 m  Warsaw telescope onto the standard system,  this
galaxy was observed on an excellent photometric night, 2006 January 26, in  V and
I bands, along with 25 Landolt standard stars which were monitored at
very different air masses, and covered a wide range of colors (-0.14  $< V-I
<$ 1.43). Because the transformation equations for each chip can have
different zero points, the standard star sample  was observed on each
chip individually. The transformation equations which were obtained for
each chip on which the galaxy was recorded (chips 2 and 3) are given in Table 3.
Lower case letters $v$ and $i$ refer to the aperture instrumental
magnitudes,  standardized to one second exposure time. Capital letters
refer to the magnitudes calibrated onto the standard system. The computed
color coefficients are consistent with those which were derived to calibrate our
earlier mosaic data from the same telescope and camera for NGC 6822 (Pietrzynski
et al. 2004), NGC 3109 (Pietrzynski et al. 2006a), NGC 55 (Pietrzynski et
al. 2006b) and WLM (Pietrzynski et al. 2007), all these galaxies being targets of the
Araucaria Project.

To correct the possible small variation of the photometric zero points in
V and I over the mosaic, the maps established by Pietrzy\'nski et al.
(2004) were used. Comparison with studies in the other Araucaria papers
revealed that these maps allow us to correct the zero-point variations
across the mosaic field in such a way that the residuals do not exceed
0.02-0.03 mag.

In order to perform an external check of our photometry, we compared it to
the photoelectric measurements obtained by
Alcaino \& Liller (1984), who established  a sequence of secondary
standards in the field of NGC 247. In our database four stars were found
to be common to this sequence. The remaining ones were either saturated or
located outside our field of view. In Table 4, the first column gives the
identification of the stars according to the Alcaino \& Liller (1984) system.
The second column is the V magnitude measured by Alcaino \& Liller. The
third column reports the V magnitude  obtained in this work.
The comparison of our V-band photometry for these objects with that of
Alcaino \& Liller demonstrates that the zero points of the two data sets
agree within 0.02 mag.

In order to calibrate the instrumental photometry obtained with the 2.2 m
and 4.0 m telescopes, linear offsets were applied to the photometric zero
points of the data since the magnitudes did not show any dependence on the
color terms. We developed an internal test to estimate the zero point
photometry accuracy of the final data set obtained by joining data of the three
telescopes. Computing the mean magnitudes of bright stars observed with
each telescope, and comparing the values with those obtained by combining
data of the three telescopes, we were able to establish that the accuracy of
the V and I zero points of the photometry is 0.02 mag and 0.01 mag,
respectively. This is borne out in the Cepheid light curves obtained
from the combined datasets, where in no case evidence for systematic offsets
between the datasets from the different cameras is observed (see Fig. 2).

\section{The Cepheid Catalog}
The photometric database was made with the 2.2 m and  4.0 m  telescopes
data. This database  was constructed using the DAOMASTER and DAOMATCH
programs (Stetson 1994). A search for variable stars with periods 
between 0.2 and 150 days was conducted using the analysis of variance algorithm
(Schwarzenberg-Czerny 1989) and FNPEAKS software (Kolaczkowski private information).
In order to distinguish Cepheids from other types of variable
stars, the same criteria used by Pietrzy\'nski et al.
(2002b) were adopted. 
All light curves showing the typical shape of Cepheid variables were
approximated by Fourier series of order less than  5  and 4 for
photometric time-series in the V and I bands, respectively. Then we
rejected all candidates whose amplitudes in the V band were smaller than 0.4
mag, which is approximately the lower limit amplitude for normal classical
Cepheids. For the variable stars passing our selection criteria, mean V
and I magnitudes were derived by integrating their light curves which had
been previously converted onto an intensity scale, and converting the
results back to the magnitude scale. The quality and phase coverage of the
light curves allowed us to obtain a statistical accuracy of the derived
mean magnitudes in the V band as typically 0.01 mag for the brightest
variables, increasing to 0.03 mag for the faintest Cepheids in our sample. Our
Cepheid catalog consists of 23 members. Table 5 presents their
identifications, equatorial coordinates, periods, V and I mean magnitudes,
and their Wesenheit index  magnitudes defined as $W_I$ = I - 1.55($<V> -
<I>$) (see Udalski et al. (1999)). Other variables in NGC 247 discovered in our search
for photometric variability and not classified as Cepheids will be studied in a 
forthcoming paper. In Fig. 3, we show the V and I light curves of a typical
Cepheid we have discovered in NGC 247, cep015 with a period of 41.4 days, as obtained
from the combined datasets. It can be seen that the different datasets match each
other very well indeed, and that the light variations of this (and the other) Cepheid
in our catalog in the V and I bands are very well determined.

The relatively small number of Cepheids detected from our images in NGC 247 is not 
surprising, given the distance to this galaxy and the fact that most of the frames
were obtained on small telescopes (there were only 9 epochs obtained on the Blanco
4-m telescope),
which restrict the detection to the bright, long-period Cepheid population
in this galaxy. On the other hand, this is the population most useful for the
distance determination, which is the primary purpose of our project, and a sample
of 23 Cepheids with a period baseline of more than 100 days is very well
suited for an accurate distance determination.

In Fig. 2, we show the locations of the NGC 247 Cepheids in the V, V-I color-magnitude
diagram (CMD), where they delineate the expected Cepheid instability strip (Chiosi
et al. 1992; Simon \& Young 1997). Three stars in Table 5 are missing in this plot 
because we could not measure their I-band light curves (objects cep001, cep002 and cep013).
While the eighteen brightest Cepheids in our sample delineate the instability strip
very well, supporting their correct identification as Cepheid variables, the two faintest
Cepheids in our sample appear too blue for their periods and seem to be placed beyond
the blue edge of the Cepheid instability strip. While this is likely to be attributed
to the relatively low accuracy of the photometric data for these two stars, it could
also mean that these objects are not Cepheids, which we consider very unlikely however
from an inspection of their light curves. In any case, as discussed in the following
section of this paper, these two variables are excluded from our distance determination
for NGC 247.

In Table 6, we report our individual V and I band observations of the
discovered Cepheids. The full Table 6
is available in the electronic edition of the journal.

\section{Period-Luminosity Relations and Distance Determination}

In Figures 4, 5 and 6 we show the PL relations in the V, I and Wesenheit bands (the latter
defined in the usual way) resulting from the data 
from Table 5. For the distance determination from these diagrams, we adopt the eighteen
Cepheids in the period range from 27.8 to 73.3 days. The variables with smaller
periods in Table 5 were excluded from the distance analysis because they might be
affected by a Malmquist bias, given that their mean magnitudes are relatively close
to the cutoff in our photometry (see Fig. 2). We also exclude the one very long-period
variable at 131.3 days because such very luminous Cepheids might not obey the
linear PL relation defined by the shorter period (less than 100 days) variables (e.g.
Freedman et al. 1992; Gieren et al. 2004). Such an effect is also predicted from
theoretical models (Bono et al. 2002).
Although the 131.3 day Cepheid in NGC 247 seems to perfectly
fit the linear extension of the PL relation in both V and I, it carries a strong
weight in the solution and we prefer not to use it in the distance calculation. 

As in the previous papers in this series, we adopt the slopes of the Cepheid PL
relations as given by the OGLE-II project for the Cepheids in the LMC (Udalski 2000).
These slopes are extremely well determined, with 1 $\sigma$ errors of 0.021, 0.014 and
0.008 in the V, I and the reddening-free Wesenheit bands, respectively.
Least-squares fits to the observed PL relations in NGC 247 from our data yield slopes of
-2.571 $\pm$ 0.261 in V, -2.685 $\pm$ 0.214 in I and -3.497 $\pm$ 0.421 in ${\rm W}_{\rm I}$,
which are all consistent with the OGLE LMC slopes within about 1 $\sigma$. This supports
our adopted procedure to use the well-measured LMC Cepheid PL relation slopes
for the distance determination of NGC 247.

Fitting the LMC Cepheid PL relation slopes to our data of the 18 adopted
Cepheids (indicated as filled circles in Figs. 4-6) yields the following equations:  \\

V = -2.775 log P + (26.766 $\pm$ 0.048)    $\sigma$ = 0.199 \\

I = -2.977 log P + (26.131 $\pm$ 0.056     $\sigma$ = 0.229 \\

${\rm W}_{\rm I}$ = -3.300 log P + (25.163 $\pm$ 0.086)   $\sigma$ = 0.355 \\

If we fit the data of {\it all} Cepheids in Table 5 (23 stars) with the same slopes,
the zero points in the above PL relations change to 26.735, 26.134 and 25.219
in V, I and the Wesenheit index, respectively. 
This shows that our distance result for NGC 247 changes
by less than 3\% if we replace our adopted sample of 18 Cepheids by the full
sample of Cepheids we detected and measured in NGC 247, in any filter. If we take
into account the small errors on the adopted LMC PL relation slopes in our calculation, 
the effect on the zero points of the NGC 247 PL relations is less tha 0.01 mag
in each band, and therefore insignificant.

Adopting, as in our previous papers, a value of 18.50 for the true distance
modulus of the LMC, a value of 0.02 mag for the foreground reddening toward NGC 247
(Schlegel et al. 1998), and the reddening law of Schlegel et al. (1998) 
( ${\rm A}_{\rm V}$ = 3.24 E(B-V), ${\rm A}_{\rm I}$ = 1.96 E(B-V)) we obtain
the following absorption-corrected  distance moduli for NGC 247 in the three
different bands:\\

$(m-M)_{0}$ (V) = 28.135 $\pm$ 0.048 \\

$(m-M)_{0}$ (I) = 27.998 $\pm$ 0.056 \\

$(m-M)_{0}$ (${\rm W}_{\rm I}$) = 27.795 $\pm$ 0.086 \\

These values suggest that in addition to the foreground reddening towards
our target galaxy, there is substantial intrinsic reddening in NGC 246 affecting
the Cepheid magnitudes in V and I. We will determine this intrinsic component
of the total reddening affecting the NGC 247 Cepheids with high accuracy
in a future study which
will provide near-infrared photometry for the Cepheids and combine it with the
present optical VI data, as we did in previous papers of the Araucaria Project
(e.g Gieren et al. 2005a). From the distance moduli derived in the V and I bands,
our current best estimate for the {\it total} mean reddening of the NGC 247 Cepheids,
including that produced in the galaxy itself,is E(B-V)=0.13 mag. This value of
the total reddening brings the true distance moduli in the V and I bands
in very good agreement with the value derived from the ${\rm W}_{\rm I}$. As our present
best determination of the true distance modulus of NGC 247, we adopt the value
27.80 $\pm$ 0.09 mag (random error) derived from the reddening-free Wesenheit magnitude.
The corresponding distance of NGC 247 is ( 3.63 $\pm$ 0.15) Mpc and places
the galaxy at about twice the distance of the other two Sculptor Group galaxies
for which our project has provided accurate Cepheid distances so far, NGC 300
(Gieren et al. 2004, 2005a) and NGC 55 (Pietrzy{\'n}ski et al. 2007, 
Gieren et al. 2008). We will discuss our distance determination for NGC 247
and estimate its total uncertainty, including systematic errors, in the following
section.

\section{Discussion}
The current distance determination to NGC 247 is the first one based on
Cepheid variables discovered in our present wide-field imaging survey. It
is affected to some extent by the well-known sources of systematic error inherent 
in such studies which have been discussed in some detail in our previous
wide-field Cepheid surveys of Araucaria galaxies mentioned in the Introduction. 
These systematic error sources include the zero point of our photometry,
possible problems with the Cepheid sample itself (overtone pulsators, Malmquist
bias, filling of the instability strip), blending of the Cepheids with objects
not resolved in the images, possible metallicity effects on the PL relation,
reddening, and the adopted distance of the LMC to which the NGC 247 distance is tied.
We will in turn discuss these factors and estimate their influence on our
current distance determination.

As discussed in section 2, we have been very careful to determine the photometric
zero points of the different sets of photometric data used in this study with
the highest possible accuracy. Our tests and discussion in section 2 lead us
to believe that the photometric zero points of the common datasets in V and I
are determined to better then $\pm$ 0.03 mag in both bands. This is in
agreement with our previous experience with datasets obtained with the same
instruments and reduced and calibrated in the same way,
which always led to photometric zero points whose accuracy
was better than 2\%.

While the Cepheid sample used in this study of about 20 stars is relatively
small as compared to our previous surveys of other (and nearer) Araucaria
galaxies, it is still large enough to expect that an inhomogeneous filling
of the instability strip is not a significant problem in the present case
of NGC 247. Indeed, the dispersion of the positions of the Cepheids in
Fig. 3 seems to indicate that the distribution of the Cepheids in the
instability strip is approximately random. This idea is supported by the fact
that the effect of sample size, or adopted cutoff period 
on the distance solution is only $\sim$ 2\% in
all bands, as discussed in the previous section. This same result also
suggests that there is no significant Malmquist bias affecting our distance
result, in agreement with the fact that even the faintest Cepheids in our
sample are still about one magnitude brighter than the faintest stars in
NGC 247 detected and measured from our images. Also, we can clearly assume
that our adopted Cepheid sample has only fundamental mode pulsators, given that
many studies have shown that overtone Cepheids are restricted to periods
smaller than about 6 days (e.g. Udalski et al. 1999; Gieren et al. 1993). 

Blending of the Cepheids in NGC 247 used for the distance determination is potentially
a more serious problem. In extrapolation of our result of the effect of blending
on the Cepheid-based distance of NGC 300 (Bresolin et al. 2005), which
was found less than 2\%, we would expect a slightly larger effect for NGC 247 due to its
larger inclination as compared to NGC 300, and the larger distance which both tend to
increase the probability of significant blending with unresolved and relatively
bright nearby stars. On the other hand, the relative effect of blending
decreases with the period of the Cepheids; longer-period Cepheids are more
luminous, and therefore the relative effect of a given unresolved companion star
on its observed flux tends to be less important. Since all Cepheids in our
sample used for the distance determination have periods longer than 27 days,
and the mean period of the sample is about 50 days, our NGC 247 sample should be
less affected in this regard than the sample in NGC 300 for which the mean period
is considerably shorter (about 30 days; Gieren et al. 2004). From
these considerations, we expect that any residual
systematic effect on the NGC 247 distance due to blending of the Cepheids
should not exceed 4\%, or 0.08 mag in the distance modulus. It should be noted
that the sign of this error is to {\it underestimate} the true distance of the
galaxy by such an amount.

The effect of metallicity on the PL relations in different bands has been
discussed in some detail in the previous papers of this series. Here we just
note that for any of the Araucaria target galaxies we have studied so far
the observed slopes of the Cepheid PL relations in the VIJK bands were
consistent with the respective slopes found in the LMC by the OGLE-II project in VI bands,
and by Persson et al. (2004) in the near-infrared bands. NGC 247 is no exception,
as evident from Figs. 4, 5 and 6, although the slope is less well determined
in the present case of NGC 247 due to the relatively small Cepheid sample
we were able to detect. The recent work of
Gieren et al. (2005b), and of Fouqu{\'e} et al. (2007) on Galactic and LMC Cepheid 
distances measured from the infrared surface brightness 
technique (Fouqu{\'e} \& Gieren 1997) has confirmed that the slope of the
Cepheid PL relation does not change significantly with metallicity even for
a solar abundance sample of Cepheid variables. The effect of a possible nonlinearity
of the PL relation at a period of 10 days discussed by Ngeow et al. (2008)
does not affect the determination of distances of galaxies in any significant way.
The question of the effect of metallicity on the zero point of the PL relation
is less clear, at the present time, and it is one of the main goals to
determine this effect by comparison of the Cepheid distances of our target
galaxies with those derived from other methods. However, the effect of
metallicity on the PL relation zero points in VIJK is certainly not dramatic
and likely less than $\sim$ 3\% (e.g. Storm et al. 2004; Sakai et al. 2004;
Macri et al. 2006).
The generally very good agreement of our Cepheid distances with those derived
from the TRGB method (e.g. Rizzi et al. 2006, 2007) for common galaxies seems 
to support the idea that the effect of metallicity on the PL zero point is
small. For NGC 247, the agreement of the TRGB distance of 27.9 $\pm$ 0.1 mag
of Davidge (2006) with our present value of 27.80 mag is clearly excellent.
Although we do not have any detailed information on the metallicity of
the young stellar population in NGC 247, it is likely more metal-rich than
that of the less massive Local Group irregular galaxies  WLM, NGC 3109 and IC 1613
for which the metallicity of young stars is about -1.0 dex (Bresolin et al. 2006,2007;
Evans et al. 2007), suggesting that Cepheid and TRGB distances do agree
well over a range of metallicities, and not only for very metal-poor galaxies.
A detailed quantitative discussion of this issue will be given in a
forthcoming paper of the Araucaria Project.

Any residual systematic effect of reddening on our present distance result adopted
from the Wesenheit PL relation should be small. In our previous Cepheid
studies, we have usually found that the distance of a galaxy derived from the
${\rm W}_{\rm I}$ magnitude PL relation agreed to within 2\% with the one
derived from our combined optical-near infrared technique (Gieren et al. 2005a,
2006, 2008a; Pietrzy{\'n}ski et al. 2006c; Soszy{\'n}ski et al. 2006). This is in principle
expected due to the reddening-free nature of the Wesenheit magnitudes. However,
the Wesenheit band PL relations for more distant galaxies tend to show an increasing 
dispersion which is in the
order of, or even surpasses the one observed in V and/or I, which is likely
due to the combined effect of photometric errors and blending with unresolved
nearby stars which  affect the Wesenheit magnitudes by increasing amounts.
In our recent
study of the WLM galaxy (Gieren et al. 2008b), we find a distance from near-infrared
data which is 0.09 mag shorter than the one we had previously obtained from
the Wesenheit index. In the case of NGC 247, there might be an effect of similar
order. Future, planned infrared photometry of the Cepheids in NGC 247 will allow
to investigate this question and provide a more accurate distance determination
to the galaxy. We estimate that the remaining systematic effect due to
reddening on the present distance modulus determination for NGC 247 should
not be larger than 4\%. Our planned infrared work on the NGC 247 Cepheids 
will definitively prove or disprove the validity of our current estimation
of the effect of reddening on the distance of NGC 247.

Finally, there is the problem of the adopted LMC distance, which has been most
recently discussed by Schaefer (2008). We do not wish to add to the discussion here,
but just mention that some recent results for the distance modulus of the LMC
using local open clusters (An et al. 2007), revised Hipparcos Cepheid parallaxes 
(van Leeuwen et al. 2007),
and the infrared surface brightness technique (Fouqu{\'e} et al. 2007) have
yielded a somewhat shorter LMC distance, with values close to 18.40 mag.
Since all galaxy distance moduli reported in previous papers of the Araucaria Project
have been tied to the same, assumed LMC distance modulus of 18.50, we will retain this
value here. Therefore,
their {\it relative} distances will not be affected should future work definitively change
the adopted 18.50 mag value of the LMC distance. 

The distance of NGC 247 seems very well determined now. Within their respective
uncertainties, the TRGB distance of 27.9 $\pm$ 0.1 mag
of Davidge (2006) and the Tully-Fisher distance determination of 27.87 $\pm$ 0.21 mag
of Karachentsev et al. (2006) agree with the present, first
distance determination of the galaxy from Cepheid variables. The new Cepheid distance
to NGC 247 reported in this paper has an uncertainty which is similar to the
TRGB measurement of the distance, but more accurate than the distance of
this spiral derived from the Tully-Fisher method.

 From the previous
discussion, we estimate that the systematic uncertainty of our present
distance determination is about 0.09 mag. This leads to a final, adopted 
value for the true distance modulus of NGC 247 of $(m-M)_{0}$ = 27.80 $\pm$ 0.09
(random) $\pm$ 0.09 (systematic) mag, corresponding to 3.63 Mpc, with a total
uncertainty of about 6\%. We expect to improve on this value, and significantly
reduce its uncertainty, with our planned near-infrared work on the NGC 247
Cepheids.

\section{Conclusions}
The main conclusions of this paper can be summarized as follows:

1. We have conducted an extensive wide-field imaging survey in the Sculptor
Group spiral galaxy NGC 247 in the optical V and I bands which has led
to the discovery of 23 Cepheid variables with periods between 17 and 131 days.
These are the first Cepheid variables with published data reported in this
galaxy. 

2. From the periods and mean magnitudes of the Cepheids derived from our photometry,
we have constructed PL relations in the V, I and ${\rm W}_{\rm I}$ bands. The
data define tight PL relations which are well fitted with the slopes of
the corresponding LMC Cepheid PL relations adopted from the OGLE-II project.
We have derived reddened distances for the 3 bands, which lead to a determination
of the total mean reddening of the NGC 247 Cepheids of E(B-V) = 0.13 mag.
With an adopted foreground reddening towards NGC 247 of 0.02 from Schlegel et
al. (1998), we determine a reddening value of 0.11 produced {\it inside} NGC 247
and affecting its Cepheid population. We adopt a final true distance modulus of
27.80 mag for NGC 247 from the reddening-free Wesenheit magnitude PL relation,
in very good agreement with the reddening-corrected values coming from the
V and I bands. The total uncertainty on this distance value, including the
contribution from systematic errors, is estimated to be $\pm$ 6\%.

3. The Cepheid distance to NGC 247 derived in this paper agrees very well,
within the respective 1 $\sigma$ uncertainties, with the recent determinations of
Davidge (2006) from the I-band TRGB method, and of Karachentsev et al. (2006)
from the Tully-Fisher method. 

4. The distance of NGC 247 of 3.63 Mpc is about twice as large as the respective
distances of two other Sculptor Group galaxies, NGC 300 (Gieren et al. 2005a)
and NGC 55 (Gieren et al. 2008a). This supports the evidence that the Sculptor Group
has a filament-like structure (Jerjen et al. 1998) with a large depth extension
in the line of sight.

\acknowledgements
JAGV acknowledges financial support from  MECESUP grant USA0108 
to the Universidad de Concepci\'on. 
JAGV, GP and WG acknowledge
financial support from the Chilean Center for Astrophysics FONDAP
15010003. WG also acknowledges support from the BASAL Centro de Astrofisica y
Tecnologias Afines (CATA) for this work.
GP acknowledges support from the Polish grant N203 002 31/046
and the FOCUS
subsidy of the Fundation for Polish Science (FNP). We are grateful
for generous amounts of observing time allotted to this project
at ESO-La Silla, CTIO and LCO and thank the respective staffs for their
expert help in conducting the observations.

\begin{deluxetable}{c c c}
\tablecaption{Camera Properties}
\tablehead{
\colhead{Telescope} & \colhead{FOV} & \colhead{SF ($^{\prime \prime} / pixel$)} \\
}
\startdata
1.3 m & 35$^\prime$ $\times$ 35$^\prime$ & 0.26  \\
2.2 m & 34$^\prime$ $\times$ 33$^\prime$ & 0.24  \\
4.0 m & 36$^\prime$ $\times$ 36$^\prime$ & 0.27  \\ 
\enddata
\end{deluxetable}

\begin{deluxetable}{c c c c c}
\tablecaption{Summary of Observations}
\tablehead{
\colhead{Telescope} & \colhead{$t_{exp}$} & \colhead{Time Base-} & \colhead{Seeing} & \colhead{Band} \\
&  \colhead{(sec)}  & \colhead{Line (yr)}  & \colhead{(arcsec)}
}
\startdata
1.3 m & 900 & 1.3 & 0.8-1.4 & V, I  \\
2.2 m & 700 & 2.3 & 0.7-1.5 & V  \\
4.0 m & 300, 720 & 2.2 & 1.0-1.7 & V  \\
4.0 m & 300, 360 & 2.2 & 1.0-1.7 & I  \\ 
\enddata
\end{deluxetable}

\begin{deluxetable}{c c}
\tablecaption{Photometric Calibration for the 1.3 m Warsaw Telescope Data}
\tablehead{
\colhead{Chip 2} & \colhead{Chip 3} \\
}
\startdata
$V=v-0.032(V-I)-2.339$&$V=v-0.030(V-I)-1.898$\\
$I=i+0.031(V-I)-2.572$&$I=i+0.037(V-I)-2.283$\\
$V-I=0.939(v-i)+0.226$&$V-I=0.936(v-i)+0.365$\\
\enddata
\end{deluxetable}

\begin{deluxetable}{c c c}
\tablecaption{Photometry Comparison}
\tablehead{
\colhead{ID} & \colhead{$V_{A\&L}$} & \colhead{V} \\
}
\startdata
J&12.58&12.585\\
M&13.47&13.490\\
N&14.25&14.287\\
O&13.83&13.841\\
\enddata
\end{deluxetable}

\begin{deluxetable}{c c c c c c c c}
\tablecaption{Catalog of Cepheid Variables in NGC 247}
\tablehead{
\colhead{ID} & \colhead{R.A.} & \colhead{Decl.} & \colhead{\textit{P}} &
\colhead{${\rm T}_{0}-2,450,000$}
& \colhead{$<V>$} & \colhead{$<I>$}  & \colhead{$<W_{\rm I}>$}\\
 & \colhead{(J2000)} & \colhead{(J2000)}  &
\colhead{(days)} & \colhead{(days)} &
\colhead{(mag)} & \colhead{(mag)} & \colhead{(mag)}
}
\startdata
cep001 & 0:47:29.60  & -20:45:28.3    & 15.949 & 2,807.3298  & 23.372 &  &
 \\
cep002 & 0:46:51.06  & -20:33:18.4    & 17.837 & 2,805.8816  &  23.357 & 
&  \\
cep003 & 0:46:59.67  & -20:39:30.0    & 17.966 & 2,576.1692  &  22.951 &
22.389 & 22.389\\
cep004 & 0:47:06.69  & -20:37:10.2    & 18.424 & 2,821.5388  &  22.830 &
22.402 & 22.402\\
cep005 & 0:47:07.59  & -20:37:51.4    & 27.821 & 2,586.0627  &  22.431 &
21.603 & 21.603\\
cep006 & 0:47:11.60  & -20:44:08.2    & 29.585 & 2,799.9832  &  22.601 &
21.726 & 21.726\\
cep007 & 0:47:03.39  & -20:44:47.7    & 29.977 & 2,582.8054  &  22.604 &
21.456 & 21.456\\
cep008 & 0:47:03.46  & -20:47:58.5    & 30.978 & 2,559.6931  &  22.609 &
21.713 & 21.713\\
cep009 & 0:47:09.93  & -20:39:28.6    & 32.114 & 2,587.5518  &  22.496 &
21.599 & 21.599\\
cep010 & 0:47:00.54  & -20:37:09.7    & 33.023 & 2,811.5498  &  22.646 &
21.819 & 21.819\\
cep011 & 0:47:10.11  & -20:48:45.1    & 33.172 & 2,558.3767  &  22.514 &
21.481 & 21.481\\
cep012 & 0:47:17.14  & -20:44:18.8    & 35.809 & 2,580.2461  &  22.145 &
21.239 & 21.239\\
cep013 & 0:47:09.62  & -20:51:38.2    & 36.192 & 2,803.8462  &  22.415 & 
&  \\
cep014 & 0:46:56.65  & -20:41:34.7    & 39.747 & 2,569.9172  &  22.129 &
21.177 & 21.177\\
cep015 & 0:46:58.44  & -20:44:02.9    & 41.393 & 2,579.1355  &  22.143 &
21.156 & 21.156\\
cep016 & 0:47:10.52  & -20:47:21.3    & 44.481 & 2,566.2730  &  22.351 &
21.186 & 21.186\\
cep017 & 0:47:03.82  & -20:41:04.0    & 48.663 & 2,583.7842  &  21.968 &
21.035 & 21.035\\
cep018 & 0:47:04.28  & -20:47:42.2    & 63.505 & 2,581.4636  &  22.047 &
20.952 & 20.952\\
cep019 & 0:47:10.19  & -20:42:11.7    & 64.889 & 2,597.1514  &  22.034 &
21.245 & 21.245\\
cep020 & 0:47:10.64  & -20:40:11.0    & 65.862 & 2,589.5732  &  21.983 &
21.114 & 21.114\\
cep021 & 0:47:04.85  & -20:44:22.6    & 69.969 & 2,595.1423  &  21.661 &
20.536 & 20.536\\
cep022 & 0:47:09.29  & -20:51:11.9    & 73.300 & 2,564.7302  &  21.850 &
20.791 & 20.791\\
cep023 & 0:47:08.91  & -20:43:21.4    &131.259 & 2,563.3052  &  21.019 &
19.855 & 19.855\\
\enddata
\end{deluxetable}

\begin{deluxetable}{ccccc}
\tablecaption{Individual V and I Observations}
\tablehead{
\colhead{object}  & \colhead{filter} &
\colhead{HJD-2450000}  & \colhead{mag}  & \colhead{$\sigma_{mag}$}\\
}
\startdata
cep001	& V &	2814.91283 & 23.616 & 0.119\\
cep001	& V &	2931.81538 & 23.508 & 0.128\\
cep001	& V &	2931.82500 & 23.644 & 0.158\\
cep001	& V &	2934.62648 & 22.706 & 0.038\\
cep001	& V &	2937.78340 & 22.939 & 0.112\\
cep001	& V &	2966.63869 & 22.640 & 0.056\\
cep001	& V &	2966.65743 & 22.721 & 0.077\\
cep001	& V &	2968.58759 & 23.074 & 0.095\\
cep001	& V &	2968.59719 & 23.021 & 0.074\\
cep001	& V &	2968.60686 & 23.086 & 0.073\\
cep001	& V &	2968.62680 & 22.936 & 0.067\\
cep001	& V &	2968.63630 & 23.052 & 0.063\\
cep001	& V &	2970.65025 & 23.190 & 0.085\\
cep001	& V &	2970.65964 & 23.140 & 0.076\\
cep001	& V &	2970.66902 & 23.229 & 0.096\\
cep001	& V &	3270.88984 & 22.811 & 0.090\\
cep001	& V &	3295.68266 & 23.965 & 0.119\\
cep001	& V &	3295.69197 & 23.927 & 0.136\\
cep001	& V &	3327.61168 & 23.930 & 0.244\\
cep001	& V &	3376.55724 & 23.822 & 0.130\\
cep001	& V &	3639.79773 & 23.201 & 0.082\\
cep001	& V &	3639.80704 & 23.283 & 0.096\\
cep001	& V &	3651.77611 & 22.852 & 0.061\\
cep001	& V &	3651.78545 & 22.851 & 0.065\\
cep001	& V &	3651.79484 & 22.807 & 0.059\\
cep001	& V &	3681.73375 & 23.472 & 0.117\\
cep001	& V &	3681.75232 & 23.540 & 0.135\\
\enddata
\tablecomments{The complete version of this table is in the electronic
edition of the Journal.  The printed edition contains only
the measurements in the V band for the Cepheid variable cep001.}

\end{deluxetable}

\begin{figure}[htb]
\vspace*{22cm}
\includegraphics{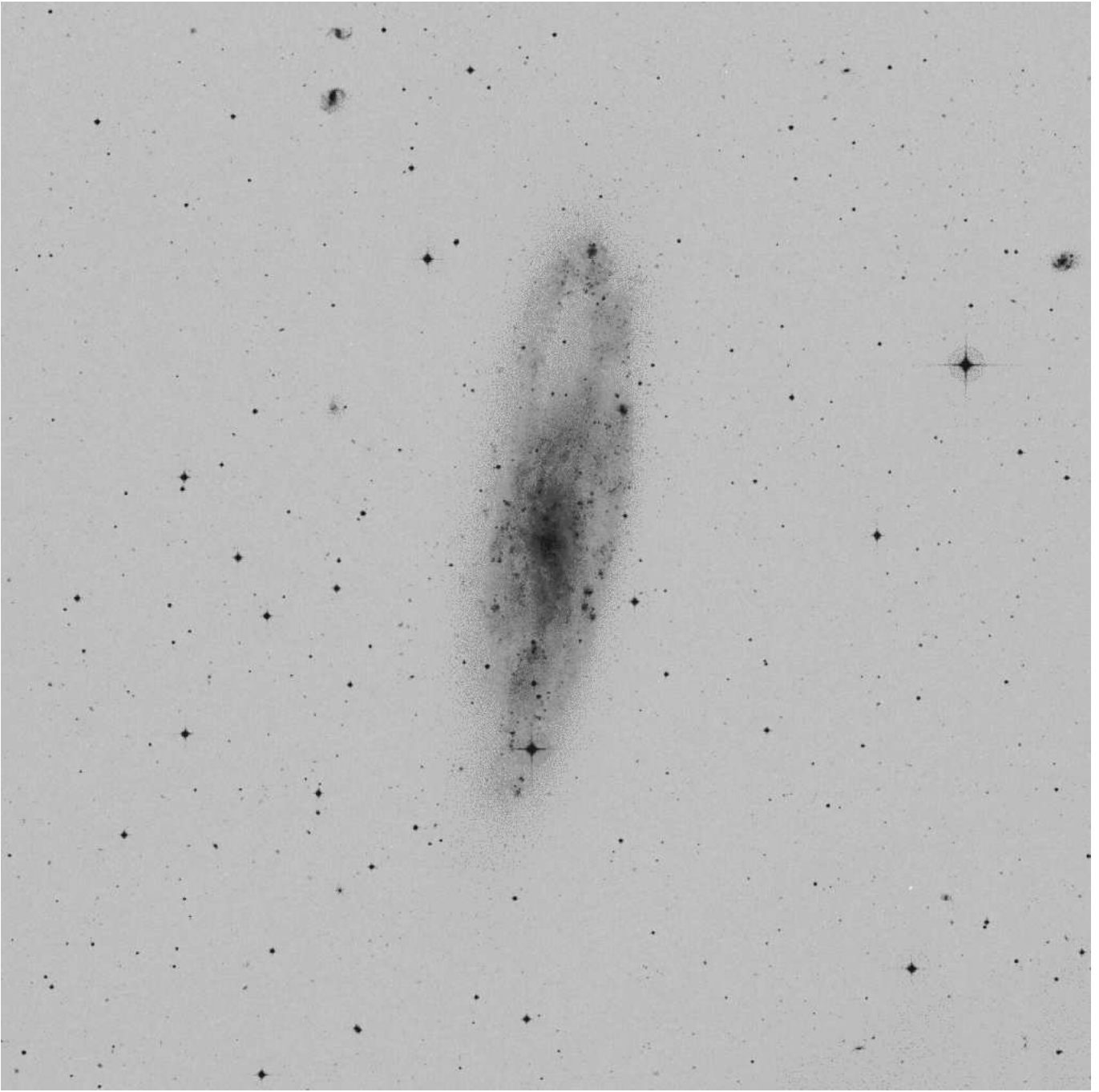}
\caption{DSS map of the observed region centered on NGC 247 galaxy.
The field of view is 35 x 35 arcmin.
}
\end{figure}

\begin{figure}[htb]
\vspace*{18cm}
\includegraphics{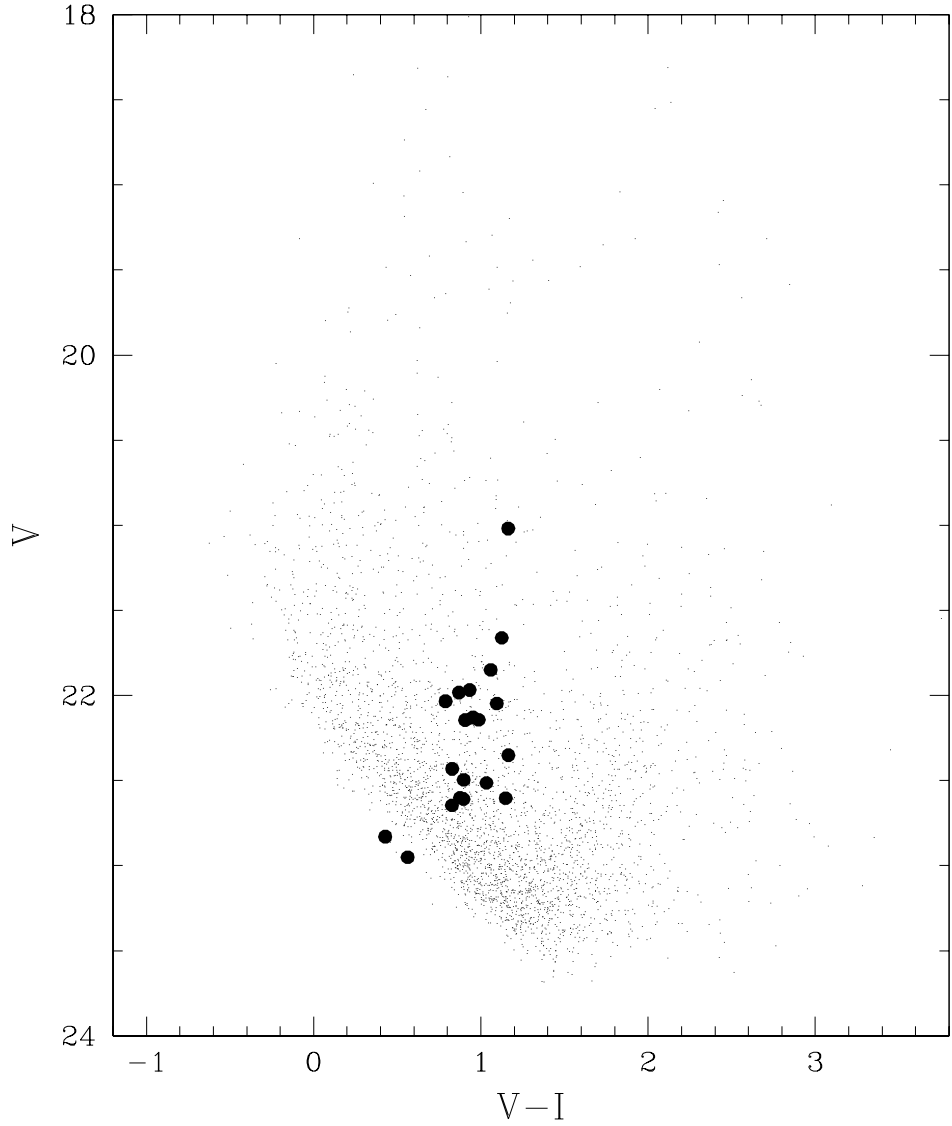}
\caption{V, V-I color-magnitude diagram for stars in NGC 247 observed with the 1.3 m
Warsaw telescope. The detected Cepheid candidates are marked with filled circles. With the
exception of the two faintest variables which are very blue for their periods,
the Cepheids populate the expected region in this diagram, supporting their
correct identification as Cepheid variables.
}
\end{figure}

\begin{figure}[htb]
\vspace*{18cm}
\includegraphics{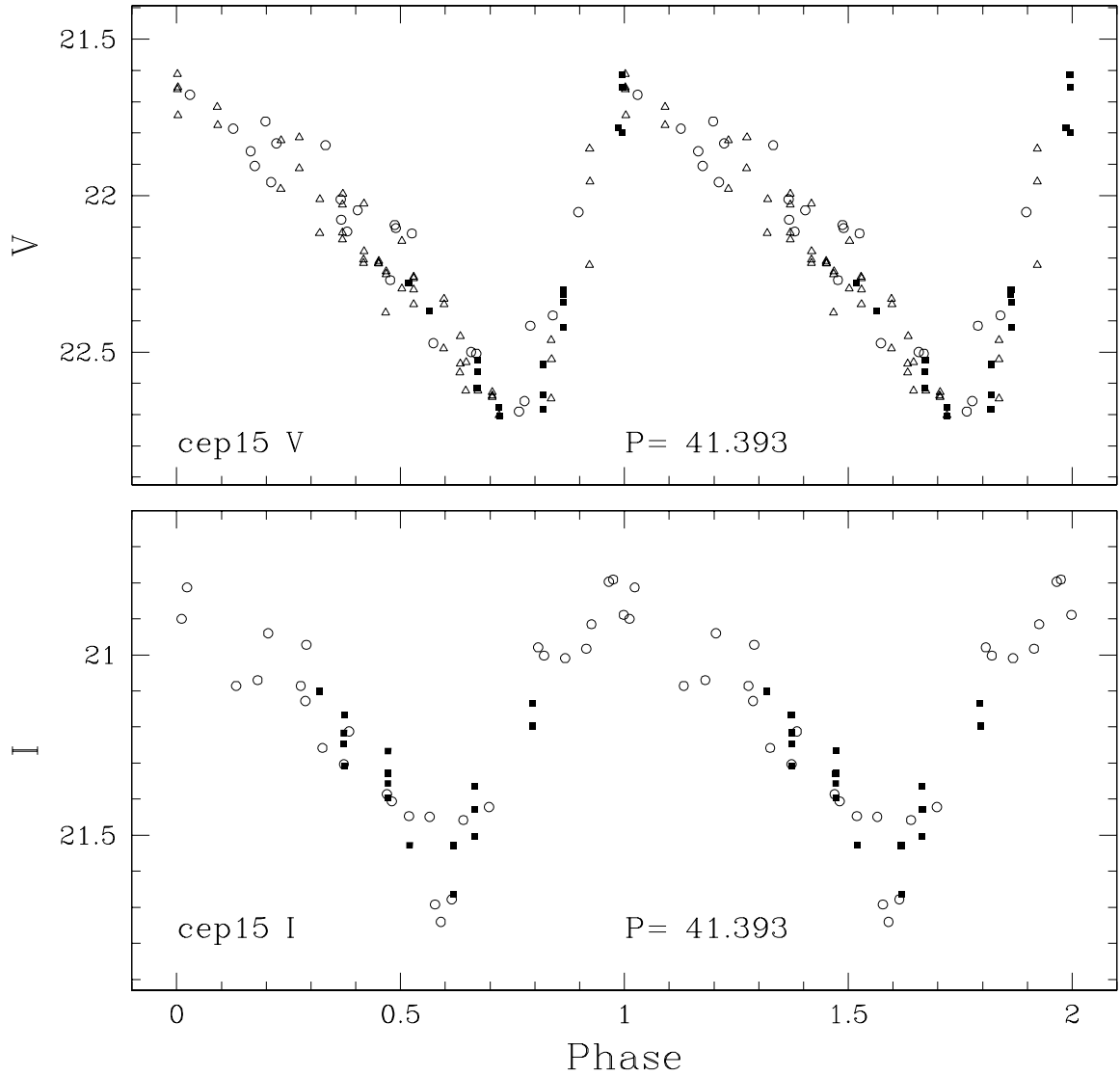}
\caption{Phased V and I band light curves for one of the Cepheids
(cep015) discovered in NGC 247. Open circles, triangles and filled 
squares correspond to the data obtained with the Warsaw 1.3 m, ESO 2.2m 
and CTI 4m telescopes,  respectively. The good agreement of the
different data sets is demonstrated.
}
\end{figure}

\begin{figure}[htb]
\vspace*{15cm}
\includegraphics{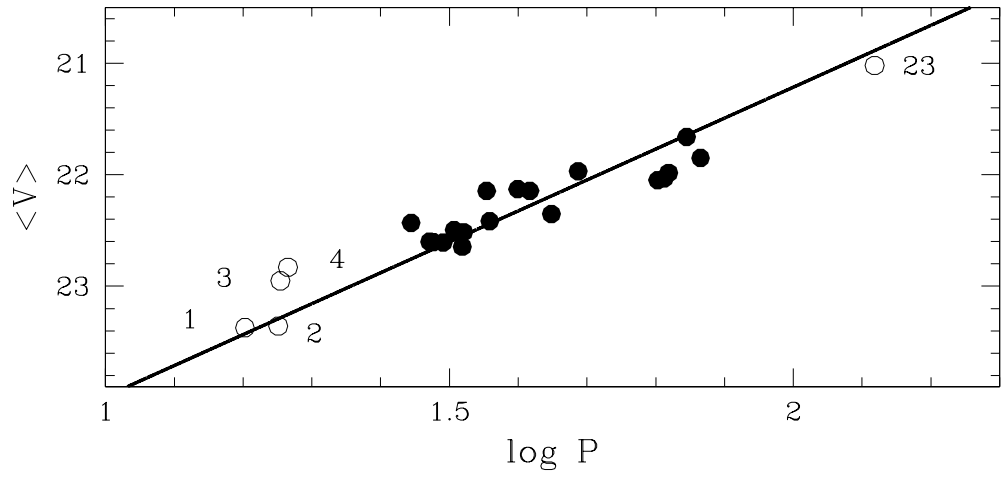}
\caption{The V-band period-luminosity relation defined by the Cepheid variables 
in NGC 247. Open circles denote stars not used for the distance
determination.
}
\end{figure}

\begin{figure}[htb]
\vspace*{15cm}
\includegraphics{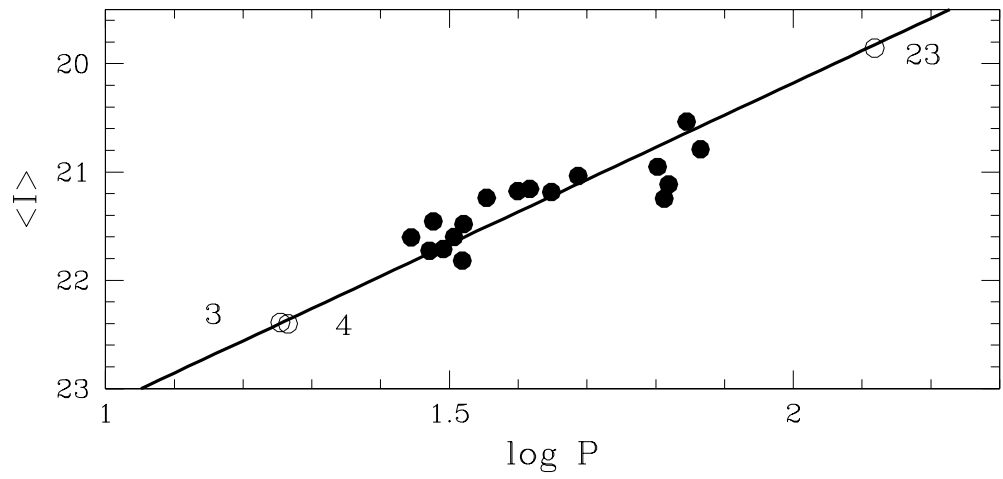}
\caption{Same as Fig. 4, for the I band.
}
\end{figure}

\begin{figure}[htb]
\vspace*{15cm}
\includegraphics{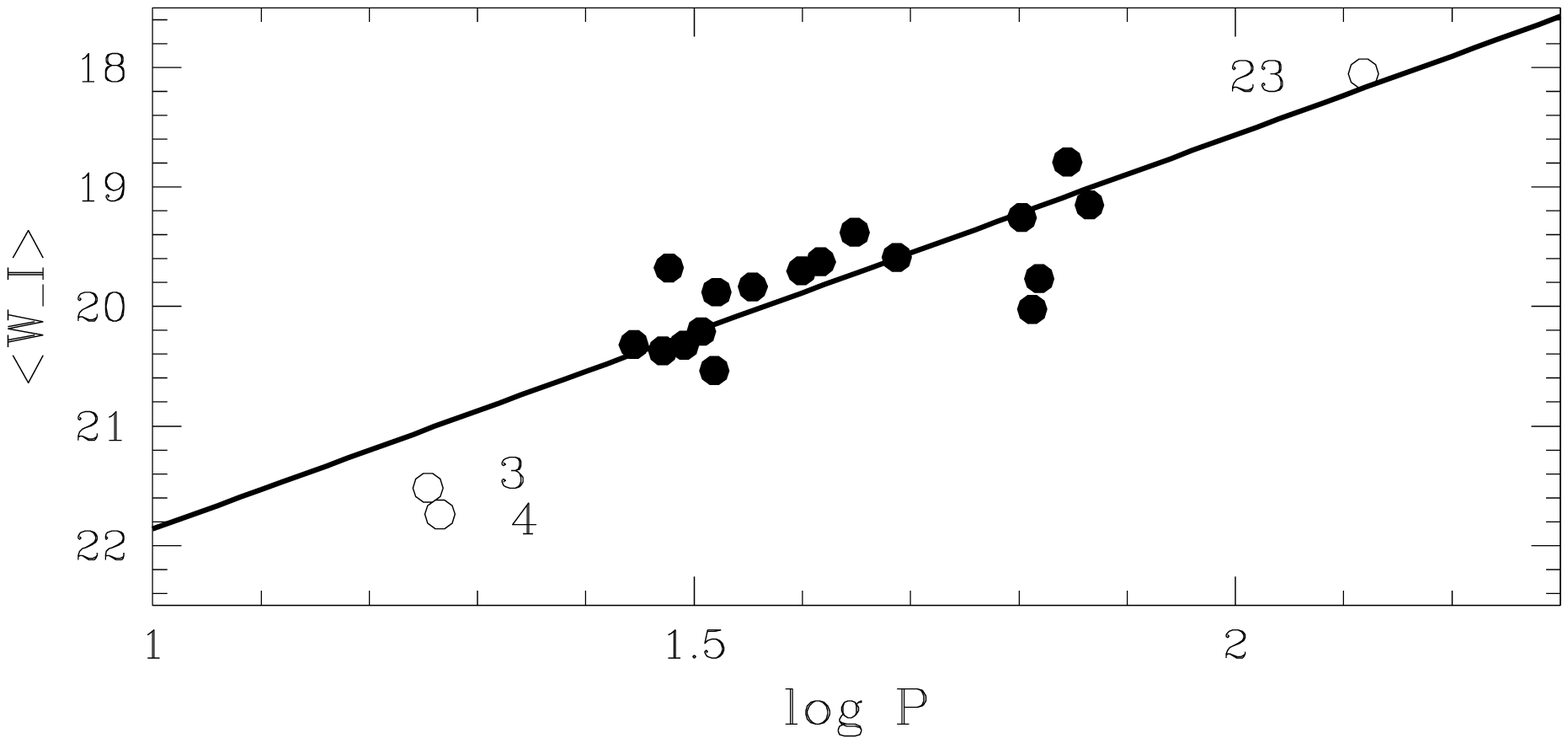}
\caption{Same as Fig. 4, for the reddening-independent (V-I) Wesenheit 
magnitudes. 
}
\end{figure}

\end{document}